\documentclass[12pt,article]{elsarticle}
\usepackage{amsmath,amssymb,amsfonts, bm, natbib}
\usepackage{dsfont}
\usepackage{epsfig}
\usepackage{graphicx}
\usepackage{slashed}
\usepackage{color}

\usepackage{caption}
\usepackage{subcaption}
\usepackage{slashed}

\usepackage{commath}
\usepackage{calc}

\newcommand{\be}{\begin{equation}}
\newcommand{\ee}{\end{equation}}

\newcommand{\bea}{\begin{eqnarray}}
\newcommand{\eea}{\end{eqnarray}}

\newcommand{\bfig}{\begin{figure}}
\newcommand{\efig}{\end{figure}}

\makeatletter
\newcommand*{\rom}[1]{\expandafter\@slowromancap\romannumeral #1@}
\makeatother

\begin{document}
\begin{frontmatter}
\title{A  model of spontaneous $CP$ breaking at low scale}
\author{Gauhar Abbas}
%\email{gauhar@prl.res.in}
\address{Theoretical Physics
Division, Physical Research Laboratory, Navrangpura, Ahmedabad 380
009, India}
\ead{gauhar@prl.res.in}

\begin{abstract}
We introduce a $CP$ symmetric model where masses of fermions are given by dimensional-5 operators, and $CP$ is spontaneously broken at TeV scale.  The unique feature of the model is that new $CP$ symmetric gauge sector coexists with new $CP$ symmetric fermions simultaneously at TeV scale.  An ultraviolet completion of the model is also proposed.  It is observed that the fine-tuning of the SM Higgs boson mass in this model is softened in a relatively small amount approximately up to $6$ TeV.  Other interesting consequences are presence of a possible dark matter candidate whose mass may be bounded from above by the SM Higgs mass. The model may also provide an explanation of recently observed flavour anomalies.
\end{abstract}
\end{frontmatter}

%\pacs{14.60.Pq, 11.10.Hi, 11.30.Hv, 12.15.Lk}

%\maketitle

$CP$-violation can distinguish matter and antimatter in an absolute and convention-independent way.  In this sense, $CP$-violation is  more fundamental ingredient of nature in comparison to $P$ or $C$ violation.  Furthermore, it is essential to explain matter-antimatter asymmetry of Universe.  However, the amount of $CP$-violation in the standard model (SM) is too small to explain matter-antimatter asymmetry of Universe.  The observation of a small $CP$-violation in the SM may be a sign that $CP$ could be a good symmetry of a larger underlying theory, and only broken by the vacuum of that theory.

There are many successful models in the literature where parity is spontaneously broken at a high scale\cite{Pati:1974yy,Mohapatra:1974hk,Senjanovic:1975rk,Mohapatra:1979ia,Abbas:2016xgj,Abbas:2016qqc,
Foot:1991bp,Silagadze:1995tr,Berezhiani:1995yi,Cui:2011wk,Cui:2012mq,Gu:2014mga,Gu:2017mkm}.  However, models of spontaneous $CP$ breaking are theoretically very different from the models based on spontaneous parity breaking in the sense that there is no gauge extension of the SM in these models\cite{Lee:1973iz,Branco:1980sz}.  In these models, spontaneous $CP$ breaking is achieved through certain terms in the scalar potential which depend on a common non-vanishing vacuum phase\cite{Branco:1980sz}.

The problematic feature of models of spontaneous $CP$-violation is the appearance of scalar-pseudoscalar mixing\cite{Deshpande:1976yp,Lavoura:1994fv}.  This gives rise to one-loop contribution to electric dipole  moments of fermions which needs a careful fine-tuning\cite{Barger:1996jc}.

In this paper, we discuss a model of spontaneous $CP$ breaking inspired by the model proposed in Ref.\cite{Lavoura:1997pq} which avoids scalar-pseudoscalar mixing.  In the model discussed in Ref.\cite{Lavoura:1997pq},  the SM gauge symmetry is extended to  $SU(3) \otimes SU(2)_L \otimes SU(2)_L^\prime \otimes U(1)_{Y^\prime}$ where $SU(2)_L^\prime $ is the $CP$ counter-part of the SM gauge group $SU(2)_L $.  The fermionic fields transformation under  $SU(3) \otimes SU(2)_L \otimes SU(2)_L^\prime \otimes U(1)_{Y^\prime}$ is given as,
\begin{eqnarray}
l_{L}={\begin{pmatrix} \nu \\ e \end{pmatrix}}_L\sim (1,2,1,-1) &,& e_R \sim (1,1,1,-2)~~,~~ \nu_{eR} \sim(1,1,1,0);\nonumber\\
q_{L} ={\begin{pmatrix} u \\ d \end{pmatrix}}_L\sim (3,2,1,\frac{1}{3}) &,& u_R \sim (3,1,1,\frac{4}{3})~~~,~~ d_R \sim(3,1,1,-\frac{2}{3});\nonumber\\
l_{L}^{\prime}={\begin{pmatrix} \nu^\prime \\ e^\prime \end{pmatrix}}_L\sim (1,1,2,-1) &,& e^\prime_R \sim (1,1,1,-2)~~,~~ \nu_{eR}^\prime \sim(1,1,1,0);\nonumber\\ 
q_{L}^\prime ={\begin{pmatrix} u^\prime \\ d^\prime \end{pmatrix}}_L\sim (3,1,2,\frac{1}{3}) &,& u^\prime_R \sim (3,1,1,\frac{4}{3})~~~,~~ d_R^\prime \sim(1,1,1,-\frac{2}{3});
\label{eq1} 
\end{eqnarray}
where $l_L$, $q_L$ are the SM doublets of fermions, and $e_R$, $\nu_{eR}$, $d_R$ and $u_R$ are singlets under the SM. $l_L^\prime$, $q_L^\prime$, $e_R^\prime$, $\nu_{eR}^\prime$, $d_R^\prime$ and $u_R^\prime$ are $CP$ counter-parts  of the SM fermions. The quantum numbers for second and third family fermions can be defined exactly as discussed above for the first family.

The transformation of fermionic fields under $CP$ can be written as,
\be
\psi_L \stackrel{\rm CP}{\rightarrow} \gamma^0 C \overline{\psi_L^\prime}^T, \
\psi_R \stackrel{\rm CP}{\rightarrow} \gamma^0 C \overline{\psi_R^\prime}^T, 
\ee
where $\psi_L$ is a doublet  of the  gauge groups $SU(2)_L$, and  $\psi^{\prime}_L$ is a doublet of the  gauge group  $SU(2)_L^\prime$.   $\psi_R$ and $\psi^{\prime}_R$ are singlets under $SU(2)_L$ and $SU(2)_L^\prime$.

In the model discussed in Ref.\cite{Lavoura:1997pq},  the Yukawa Lagrangian, for instance for electron and its $CP$ counter-part,  can be written as,
\be
{\mathcal{L}}_{Y} = \Gamma  \bar{l_L}  \varphi_L  e_R +  \Gamma^* \bar{l^{\prime}_{L}}  \varphi_L^\prime e^{\prime}_{R} + {\rm H.c.},
\ee
where $\Gamma$ is $3 \times 3$ matrix in family space, and $\varphi_L$ is the scalar doublet Higgs field charged under the SM gauge group $SU(2)_L$ and singlet under the gauge group  $SU(2)_L^\prime$.  Similarly, the scalar doublet Higgs field $\varphi_L^\prime$ is charged under the gauge group $SU(2)_L^\prime$ and singlet under the SM gauge group $SU(2)_L$ .

The Large Hadron Collider (LHC) has not found these new fermions at TeV scale yet.   Hence, mass of the lightest charged new fermion should be at least at TeV.  The mass of $CP$ counter-part of the electron is $m_{e^\prime} = m_e  \langle \varphi_L^\prime \rangle /  \langle \varphi_L \rangle$ where $m_e$ is mass of the electron, $\langle \varphi_L \rangle=246$ GeV is the vacuum expectation value (VEV) of the SM Higgs field, and $\langle \varphi_L^\prime \rangle$ is VEV of the Higgs field which is $CP$ counter-part of the SM Higgs field.  Now, for electron mass $m_e = 0.511$ MeV, and for instance $\langle \varphi_L^\prime \rangle = 5 \times 10^8$ GeV, the mass of the $e^\prime$ fermion is $1038.65$ GeV which could be searched at the LHC.  Moreover, there are interaction terms between the SM fermions and their $CP$ counter-parts.  These interaction terms may bring down scale of $\langle \varphi_L^\prime \rangle$ slightly.  However, since we need to recover small masses of the SM neutrinos ($\approx 10^{-12}$ GeV) without fine-tuning of the neutrino Yukawa couplings, we again require a very high scale of order $10^7-10^8$ GeV or so for $\langle \varphi_L^\prime \rangle$.  In Ref. \cite{Lavoura:1997pq}, neutrinos are treated as massless. 

The unusually high value of  $\langle \varphi_L^\prime \rangle$ increases the masses of gauge bosons corresponding to the gauge group  $SU(2)_L^\prime$ to $10^7-10^8$ GeV or so, thus creating a disparity of scale in the gauge sector of the model.  Thus, being so heavy, the new gauge sector is practically inaccessible to any experiment in near future.

This problem also occurs in models based on mirror fermions and mirror symmetries\cite{Abbas:2016xgj,Chakdar:2013tca,Gu:2012in}, and  is elegantly solved in Refs.\cite{Abbas:2016qqc,Abbas:2017hzw}.  The model presented in this work is in fact inspired by the models discussed in Refs.\cite{Abbas:2016qqc,Lavoura:1997pq,Abbas:2017hzw}.

The spontaneous symmetry breaking (SSB)  in  the model follows the pattern:
\begin{equation}
SU(2)_L\otimes SU(2)_L^\prime \otimes U(1)_{Y^{\prime}} \to SU(2)_L\otimes U(1)_{Y} \to U(1)_{EM}.
\end{equation}
For achieving the SSB, we introduce two Higgs doublets which transform in the following way under $SU(3)_c \otimes SU(2)_L \otimes SU(2)_L^\prime  \otimes U(1)_{Y^\prime} $:
\begin{eqnarray}
&&\varphi_L = {\begin{pmatrix} \varphi^+ \\ \varphi^0 \end{pmatrix}}_L\sim(1,2,1,1),~\varphi_L^\prime= {\begin{pmatrix} \varphi^{\prime +} \\ \varphi^{\prime 0} \end{pmatrix}}_L\sim(1,1,2,1),
 \end{eqnarray} 
and behave under $CP$ as follows:
\begin{eqnarray}
&&\varphi_L \longleftrightarrow \varphi_L^{\prime *}
\end{eqnarray} 
Since our aim is to have new gauge sector and new fermionic sector at TeV scale simultaneously, we add two real scalar singlets  to provide masses to fermions which will be explained in the following discussion.  The quantum numbers of singlet scalar fields under $SU(3)_c \otimes SU(2)_L \otimes SU(2)_L^\prime  \otimes U(1)_{Y^\prime} $ are,
\begin{eqnarray}
 \chi:(1,1,1,0),~ \chi^\prime:(1,1,1,0),
 \end{eqnarray} 
and they transform under $CP$  as,
\begin{eqnarray}
\chi \longleftrightarrow \chi^{\prime}.
\end{eqnarray}
For lowering down the scale of spontaneous $CP$ breaking such that new gauge bosons and fermions $\psi^\prime$ are simultaneously at TeV scale, we impose a pair of discrete symmetries  $\mathcal{Z}_2$ and  $\mathcal{Z}_2^\prime$ on the fermionic fields $\psi_L$, $\psi_L^\prime$ and scalar singlets $\chi$, $\chi^\prime$ as shown in Table \ref{tab1}.  All other fields transform trivially under these symmetries. 
\begin{table}[h]
\begin{center}
\begin{tabular}{|c|c|c|}
  \hline
  Fields             &        $\mathcal{Z}_2$                    & $\mathcal{Z}_2^\prime$        \\
  \hline
  $\psi_L$                 &   +  &     -                                    \\
  $\chi$                        & +  &      -                                                   \\
  $ \psi_L^\prime$     & -  &   +                                              \\
  $\chi^\prime$           & - &      +            \\
  \hline
     \end{tabular}
\end{center}
\caption{The charges of fermionic and singlet scalar fields under $\mathcal{Z}_2$ and  $\mathcal{Z}_2^\prime$  symmetries.}
 \label{tab1}
\end{table}   
 
The Yukawa Lagrangian is forbidden by the discrete symmetries $\mathcal{Z}_2$ and  $\mathcal{Z}_2^\prime$, and masses of the first family leptons are given by the dimension-5 operator as,
\bea
\label{mass1}
{\mathcal{L}}_{mass} &=& \dfrac{1}{\Lambda} \left[  \Gamma_1 \bar{l_L}   \varphi_L \chi e_R+  \Gamma_1^{* } \bar{l^{\prime}_{L}}    \varphi_L^\prime \chi^\prime e_R^\prime + \Gamma_2  \bar{l_L}  \tilde{\varphi}_L  \chi   \nu_{e_R}  + \Gamma_2^{ *} \bar{l^{\prime}_{L}}  \tilde{\varphi}_L^\prime  \chi^\prime \nu^{\prime}_{e_R} \right]  \\ \nonumber 
&+&  \dfrac{1}{\Lambda} \left[  \rho~ \bar{l_L}    \varphi_L \chi  e_R^\prime + \rho^*~ \bar{l_L}^\prime  \varphi_L^\prime \chi^\prime  e_R + \kappa \bar \ell_L \tilde \varphi_L \chi \nu_{eR}^\prime + \kappa^*  \bar{l_L}^\prime \tilde \varphi_L^\prime \chi^\prime \nu_{eR}   \right]  +  {\rm H.c.},
\eea
where  $\tilde{\varphi}_{L} = i \tau_2 \varphi_L^*$, $\tilde{\varphi}_{L}^\prime = i \tau_2 \varphi_L^{\prime *}$, $\tau_2$ is the second Pauli matrix, and $\Gamma_i$ ($i=1,2$), $\rho$ and $\kappa$ are $3 \times 3$ matrices in family space.  A similar Lagrangian can be written for quarks of the first family and fermions of other families in general.

 \begin{figure}[t!]
 \begin{center}
  \includegraphics[scale=0.5]{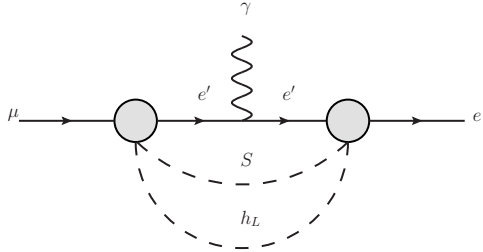}
    \caption{The additional new contribution to the process $\mu \rightarrow e \gamma$ where $h_L$ denotes the SM Higgs boson, $S$ is the scalar particle corresponding to the field $\chi$, and blob represents an effective interaction.}
    \label{mueg}
     \end{center}
\end{figure}
Since $\rho$ is a $3 \times 3$ matrix in family space, this  Lagrangian gives rise to flavour mixing in family space of fermions resulting in an additional contribution to flavour violating processes like $\mu \rightarrow e \gamma$. This additional new contribution to this process, as shown in Fig.\ref{mueg}, effectively occurs at two loops, and hence highly suppressed in this model.

The above Lagrangian provides one of the main features of the model. For keeping the mass of the gauge bosons corresponding to the gauge group $SU(2)_L^\prime$ at TeV scale, the VEV of the  field $\varphi_L^\prime$  should be such that $ \langle \varphi_L^\prime \rangle  >>    \langle \varphi_L \rangle =246$ GeV.   For instance, if VEV of the field $\varphi_L^\prime$ is such that $\langle \varphi_L^\prime \rangle = 1$ TeV which is much larger than the SM Higgs field VEV $ \langle \varphi_L \rangle$, the masses of gauge bosons corresponding to the gauge group $SU(2)_L^\prime$ should be around $1$ TeV.   Now, mass of the $CP$ counter-part of the electron is 
$m_{e^\prime} = m_e  \langle \varphi_L^\prime \rangle  \langle \chi^\prime \rangle  / \langle \varphi_L \rangle \langle \chi \rangle$.  Thus, for having the lightest new charged lepton $e^\prime$ at TeV scale, the pattern of the  spontaneous breaking of $CP$ should be such that  $ \langle \chi^\prime \rangle >>   \langle \varphi_L^\prime \rangle  >>    \langle \varphi_L \rangle$ and $ \langle \chi^\prime \rangle >> \langle \chi \rangle $.  For illustration, $ \langle \varphi_L \rangle=246$ GeV, $\langle \varphi_L^\prime \rangle = 1$ TeV, $\langle \chi \rangle =100 $ GeV, electron mass $m_e = 0.511$ MeV and $  \langle \chi^\prime \rangle = 5 \times 10^7$ GeV, the mass of the lightest new charged lepton $e^\prime$ is $1038.62$ GeV. The scale of the VEV $  \langle \chi^\prime \rangle$ may be lower when terms having interactions among the SM fermions and their $CP$ counter parts are included in the masses of fermions.

The Majorana mass term for neutrinos is given as,
\bea
\label{mass2}
{\mathcal{L}}^{\nu}_{Majorana} & =& \dfrac{1}{\Lambda} \left[ c_1~  \bar{l_{L}^c}    \tilde{\varphi}_{L}^* \tilde{\varphi}_L^\dagger  l_L  + c_1^*~ \bar{l_{L}^{\prime c}}    \tilde{\varphi}_{L}^{\prime *} \tilde{\varphi}_L^{\prime \dagger}  l_L^\prime  \right]  \\ \nonumber
&+& \dfrac{1}{\Lambda} \left[ c_2~  \bar{l_{L}^c}    \tilde{\varphi}_{L}^* \tilde{\varphi}_L^{\prime \dagger}  l_L^\prime  + c_2^*~ \bar{l_{L}^{\prime c}}    \tilde{\varphi}_{L}^{\prime *} \tilde{\varphi}_L^{ \dagger}  l_L  \right]  \\ \nonumber
& +&   M  \nu^{T}_R C^{-1}  \nu_R + M^*   \nu^{\prime T}_R C^{-1} \nu^\prime_R 
+ c_3 \nu_R^T C^{-1} \nu_R^\prime   + {\rm H.c.}
\eea
We observe that this model has an explanation  for small neutrino masses using type-\rom{1} see-saw mechanism.

The most general scalar potential of the model can be written as,
\begin{eqnarray}
V &=& -\mu_L^2  \varphi_L^\dagger \varphi_L   -\mu_L^{\prime 2}  \varphi_L^{\prime  \dagger} \varphi_L^\prime    - \mu_\chi^2  \chi^2 - \mu_{\chi^\prime}^{ 2} \chi^{\prime 2}  
+  \lambda_1 \Bigl(   (\varphi_L^\dagger \varphi_L)^2 +   (\varphi_L^{\prime \dagger} \varphi_L^\prime)^2  \Bigr)  \\ \nonumber
&+& \lambda_2  \varphi_L^\dagger \varphi_L \varphi_L^{\prime \dagger} \varphi_L^\prime   
  + \lambda_3 \Bigl(  \chi^4 + \chi^{\prime 4}  \Bigr) + \lambda_4  \chi^2 \chi^{\prime 2} 
+  \lambda_5 \Bigl(   \varphi_L^\dagger \varphi_L \chi^2 +   \varphi_L^{\prime \dagger} \varphi_L^\prime \chi^{\prime 2}  \Bigr) \\ \nonumber
&+& \lambda_{6} \Bigl(   \varphi_L^\dagger \varphi_L \chi^{\prime 2} +   \varphi_L^{\prime \dagger} \varphi_L^\prime \chi^{ 2}  \Bigr). 
\end{eqnarray}
It should be noted that we have introduced soft $CP$ breaking terms in the scalar potential which are essential to provide spontaneous $CP$ breaking  such that  $ \langle \chi^\prime \rangle = \omega^\prime/\sqrt{2} >> \langle \varphi_L^\prime \rangle=v_L^\prime /\sqrt{2}  >>    \langle \varphi_L \rangle=v_L/\sqrt{2}$ and $ \langle \chi^\prime \rangle >> \langle \chi \rangle=\omega/\sqrt{2} $.  As discussed in Ref.\cite{Lavoura:1997pq}, this model solves the strong $CP$ problem naturally. This conclusion still holds even after introducing real singlet scalar fields.   

 \begin{figure}[t!]
 \begin{center}
  \includegraphics[scale=0.5]{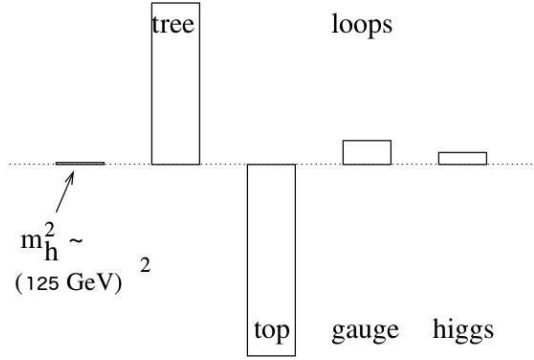}
    \caption{The fine tuning required to recover the 125 GeV
Higgs mass in the SM with a cut-off scale 6 TeV.}
    \label{fine-t}
     \end{center}
\end{figure}

This model mitigates  the fine-tuning of the SM Higgs boson mass in a relatively small amount approximately up to $6$ TeV by avoiding the  one-loop contribution to the SM Higgs boson mass due to the top quark.  For being on a concrete ground, let us assume that the SM is valid up to a cutoff scale of $6$ TeV.  Then, the main three quadratic contribution at one loop to the mass of the SM Higgs boson at the scale of  $6$ TeV  are $-3 y_t^2 \Lambda^2/ 8 \pi^2  \approx 87.5 (125)^2 \rm{GeV}^2$ from the top quark loop, $ g^2 \Lambda^2/ 16 \pi^2 \approx 11.3 (125)^2 \rm{GeV}^2$ from the gauge loop, and  $ \lambda \Lambda^2/ 16 \pi^2 \approx 5.76 (125)^2 \rm{GeV}^2$ from the Higgs loop\cite{Schmaltz:2002wx}.  Thus, the approximate mass of the SM Higgs boson is $m_h^2 \approx  m_{tree}^2 - \left( 87.5 - 11.3 - 5.76  \right) (125 \rm{GeV})^2$. This is depicted in Fig.\ref{fine-t}.  In order to recover the $125$ GeV SM Higgs mass, approximately $1\%$ fine-tuning is required. In the model discussed in this paper, one loop contribution due to top quark is absent.  Moreover, the LHC data is showing that this discovered Higgs boson is behaving like the SM Higgs boson.  Hence, one loop contributions to its mass from the scalar fields $\varphi_L^\prime$, $\chi$ and $\chi^\prime$ are expected to be small.  Therefore, we can safely take the one loop contribution to the mass of the SM Higgs in  this model at the scale of $6$ TeV to be $m_{h_L}^2 \approx m_{tree}^2 - \left( - 11.3 - 5.76  \right) (125 \rm{GeV})^2$.  Thus, we see that there is relatively less fine-tuning of the SM Higgs boson mass in this model at the scale of $6$ TeV.  Moreover, it is possible that there is a further cancellation of the above quadratic divergences by the one loop contributions from the scalar fields $\varphi_L^\prime$, $\chi$ and $\chi^\prime$.  This will be studied in future.

We remark that for cancelling the quadratic divergence of the SM Higgs boson in the supersymmetric framework, one needs a supersymmetric particle in the vicinity of the discovered Higgs boson.  However, there is no sign of such a particle in the LHC run 1 or 2.  Besides this, the decay $B_s^0 \rightarrow \mu^+ \mu^-$ which is particularly sensitive to supersymmetry, does not provide any evidence of such a particle too\cite{Aaij:2012nna,CMS:2014xfa}.  Hence, any alternative idea to deal with the fine tuning of the SM Higgs mass is worth exploring.
 \begin{figure}[t!]
 \begin{center}
  \includegraphics[scale=0.7]{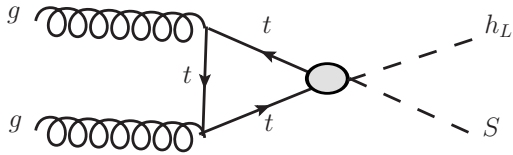}
    \caption{The production of the scalar singlet S with the SM Higgs boson $h_L$ in the gluon fusion at the LHC.  Here, blob shows an effective interaction.}
    \label{hLS}
     \end{center}
\end{figure}

It should be noted that physical particles corresponding to the  singlet scalar fields $\chi$ and $\chi^\prime$ are a mixture of  scalar fields $\chi$, $\chi^\prime$, $\varphi_L$ and $\varphi_L^\prime$ after the SSB.  The lighter physical scalar singlet particle $S$ which can be mapped onto  the  singlet scalar field $\chi$  could be a possible dark matter candidate if its mass is less than mass of the discovered SM Higgs boson (125 GeV).  The reason is that being singlet under the whole gauge symmetry of the model, it can only interact with fermions and the SM Higgs field through the Eq.(\ref{mass1}), and other scalars of the model through couplings given in the scalar potential. We need to assume that scalar particles corresponding to scalar fields  $\varphi_L^\prime$ and $\chi^\prime$ are heavier than the scalar particle corresponding to the  singlet scalar field $\chi$.  Hence, the decay of the scalar particle, $S$,  corresponding to the  singlet scalar field $\chi$, which is a mixture of  scalar fields $\chi$, $\chi^\prime$, $\varphi_L$ and $\varphi_L^\prime$,  may occur through its interaction with fermions and the SM Higgs boson given by the Eq.(\ref{mass1}). Thus if its mass is lighter than the mass of the discovered SM Higgs boson, its decay to any final states through the Eq.(\ref{mass1}) is forbidden by kinematics, and this particle can decay only at loop level.  Thus, there is an inbuilt upper bound on the mass of the possible dark matter candidate in this model.  This particle is testable at the LHC in the process of gluon fusion $gg \rightarrow h_L S$ as shown in Fig.\ref{hLS}.  The physical scalar singlet particle $S^\prime$ which can be mapped onto  the  singlet scalar field $\chi^\prime$, being the heaviest particle, is expected to decay into  lighter particles.  Hence, it is less probable a candidate for dark matter.  However, a thorough  investigation is required in future.

Besides this, the SM right handed neutrinos  and new neutrinos $\nu^\prime$ may be dark matter candidates.  However, these neutrinos also interact with gauge sector.  Hence, they are allowed to decay via a loop having a charged gauge boson and a charged fermion to a neutrino and a photon: $\nu_i \rightarrow \nu_j \gamma$ where $\nu_{i,j}$ is either the SM right handed neutrinos or new neutrinos $\nu^\prime$, and sub-script shows the conversion of one type neutrino to other type.  Because of mixing of the SM fermions and their $CP$ counter-parts, there will be additional contribution to this process having the SM and new $\psi^\prime$ fermions in the loop.  Whether neutrinos are dark matter candidate, will be determined by the rate of the process $\nu_i \rightarrow \nu_j \gamma$.

The gauge interactions of the scalar fields are given by  the following Lagrangian:
\begin{eqnarray}
{\cal L}_{GS}& =& \left({\cal D}_{\mu,L}\varphi_L \right)^\dagger\left({\cal D}^{\mu}_L \varphi_L \right)
+  \left({\cal D}_{\mu,L}^\prime \varphi_L^\prime \right)^\dagger\left({\cal D}^{\prime \mu}_L \varphi_L^\prime \right) ,
\label{ktl}
\end{eqnarray}
where, the covariant derivatives are,
\begin{eqnarray}
\mathcal{D}_{\mu,L} (\mathcal{D}_{\mu,L}^\prime) &=& \partial_\mu + i g  \dfrac{\tau_a}{2} \mathcal{W}^a_{\mu,L} (\mathcal{W}^{\prime a}_{\mu,L}) +ig^\prime \frac{Y^\prime}{2} B_\mu,
\end{eqnarray}
where, $\tau_a$'s are the Pauli matrices, and $g$ corresponds to the common coupling of the gauge groups $SU(2)_L $ and $ SU(2)_L^\prime$.  The coupling constant of the gauge group $U(1)_{Y^\prime}$ is $g^\prime$.
 
The  masses of charged gauge bosons after the SSB are given by,
\begin{equation}
M_{W^\pm_{L}}~=~\frac{1}{2}g v_L,~~M_{ W^{\prime \pm}_{L}}~=~\frac{1}{2} g v_L^\prime.
\end{equation}  

The mass matrix of the neutral gauge bosons in the basis ($W_L^3,~ W_L^{\prime 3},~B$) can be written as,
\begin{equation}
M=\frac{1}{4}{\begin{pmatrix} g^2v_L^2 & 0 & -gg^\prime v_L^2 \\ 0 & g^2  v_L^{\prime 2} & -gg^\prime  v_L^{\prime 2} \\ -gg^\prime v_L^2 &  -g g^\prime v_L^{\prime 2} & g^{\prime 2}(v_L^2+ v_L^{\prime 2} ) \end{pmatrix}}.
\label{MZ}
\end{equation}

The  weak eigenstates of neutral gauge bosons  ($W_L^3, W_L^{\prime 3},~B$) can be converted into  the physical mass eigenstates ($Z_L,~Z_L^\prime,~\gamma$) through an  orthogonal transformation $\mathcal{T}$ given as,
\begin{equation}
{\begin{pmatrix} W_L^3 \\  W_L^{\prime 3} \\ B \end{pmatrix}}=\mathcal{T} {\begin{pmatrix} Z_L \\  Z_L^\prime \\ \gamma \end{pmatrix}}.
\label{trans}
\end{equation}

The masses of the physical neutral gauge bosons are given as,
\begin{equation}
M_{Z_L}^2= \frac{1}{4}v_L^2g^2\frac{g^2+2g^{\prime 2}}{g^2+g^{\prime 2}}\left[1-\frac{g^{\prime 4}}{\left(g^2+g^{\prime 2}\right)^{ {2}}}\epsilon\right],
M_{ Z_L^\prime}^2= \frac{1}{4} v_L^{\prime 2} \left(g^2+g^{\prime 2}\right)\left[1+\frac{g^{\prime 4}}{\left(g^2+g^{\prime 2}\right)^{ {2}}}\epsilon\right],
\label{MZ2}
\end{equation}
where  $\epsilon = v_L^2/ v_L^{\prime 2}$,  and  terms of order ${\cal O}(\epsilon^2)$  are ignored since  $v_L^\prime >> v_L$.

We can  parametrize the orthogonal matrix $\mathcal{T}$  in Eq.(\ref{trans}) in terms of the mixing angle $\theta_{W_L}$  given as,
\begin{eqnarray}
{\rm cos}^2\theta_{W_L} = \left(\frac{M_{W_L}^2}{M_{Z_L}^2}\right)_{\epsilon=0}=\frac{g^2+g^{\prime2}}{g^2+2g^{\prime 2}}.
\end{eqnarray}

Thus, the transformation matrix $\mathcal{T}$ is given as,
\begin{equation}
\mathcal{T} = {\begin{pmatrix}
-{\rm cos}\theta_{W_L} & -  \dfrac{{\sqrt{\rm cos 2 \theta_{W_L}}} {\rm tan}^2 \theta_{W_L}}{{\rm cos} \theta_{W_L}} \epsilon & {\rm sin} \theta_{W_L} \\
{\rm sin} \theta_{W_L} {\rm tan} \theta_{W_L} \left[ 1+\frac{{\rm cos} 2 \theta_{W_L}}{{\rm cos}^4\theta_{W_L}} \epsilon\right] & -\dfrac{{\sqrt{\rm cos 2 \theta_{W_L}}}}{{\rm cos} \theta_{W_L} } \left[1-{\rm tan}^4 \theta_{W_L} \epsilon \right] & {\rm sin}\theta_{W_L} \\
{\rm sin}\theta_{W_L} \dfrac{{\sqrt{\rm cos 2 \theta_{W_L}}}}{{\rm cos} \theta_{W_L} } \left[ 1-\frac{{\rm tan}^2 \theta_{W_L}}{{\rm cos}\theta_{W_L}} \epsilon\right] & {\rm tan} \theta_{W_L} \left[1+{\rm tan}^2 \theta_{W_L} \frac{{\rm cos} 2 \theta_{W_L}}{{\rm cos}^2\theta_{W_L}} \epsilon \right] &  \sqrt{\rm cos 2 \theta_{W_L}}
\end{pmatrix}}.
\end{equation}
 It should be noted that there are further terms of order $\epsilon^2$ in the transformation matrix $\mathcal{T}$, however the third column of that matrix is unchanged by those further terms.

The following relations are obtained between couplings of the gauge symmetries of the model and electric charge:
\begin{equation}
g=\frac{e}{{\rm sin}\theta_{W_L}},~~g^\prime=\frac{e}{\sqrt{\rm cos 2 \theta_{W_L}}},~~
\frac{1}{e^2}=\frac{2}{g^2}+\frac{1}{g^{\prime 2}}.
\end{equation}

Eqs.(\ref{mass1}) and (\ref{mass2})  provide masses of fermions.  For instance, we can write the Lagrangian for the down type quark and its $CP$ counter-part as,
\begin{eqnarray}
{\cal L}_d &=& \dfrac{1}{\Lambda} \left(\Gamma_d \bar q_L \varphi_L d_R \chi  + \Gamma_d^* \bar { q}^{\prime }_L  \varphi_L^\prime d^{\prime}_R \chi^\prime \right) +\dfrac{1}{\Lambda} \left[  \rho_d~ \bar{q_L}    \varphi_L \chi  d_R^{\prime} + \rho_d^*~ \bar{q_L}^{\prime } \varphi_L^\prime  \chi^\prime  d_R \right]  + {\rm H.c.}\nonumber\\
&=& {\begin{pmatrix} \bar d_L & \bar { d}^{\prime}_L \end{pmatrix}}{\begin{pmatrix}\frac{\Gamma_d v_L \omega}{2 \Lambda} & \frac{\rho_d v_L \omega }{2 \Lambda} \\ \dfrac{\rho_d^* v_L^\prime \omega^\prime}{2 \Lambda}  & \dfrac{\Gamma_d^* v_L^\prime \omega^\prime}{2 \Lambda}\end{pmatrix}}{ {\begin{pmatrix} d_R \\  { d}^{\prime}_R \end{pmatrix}}}~+~{\rm H.c.}.
\end{eqnarray} 
The mass matrix in above equation is in general $6 \times 6$.

The bi-diagonalization of the mass matrices can be achieved as discussed in Ref. \cite{Lavoura:1997pq}. The following transformations should be used:
\be
\label{bidiag}
\left( \begin{array}{c} u_L \\ u_L^\prime \end{array} \right)
=
\left( \begin{array}{c} X_{u} \\ Y_{u} \end{array} \right)
\left( \begin{array}{c} u_L \\ u_L^\prime \end{array} \right) \ \mbox{\rm and} \
\left( \begin{array}{c} d_L \\ d_L^\prime \end{array} \right)
=
\left( \begin{array}{c} X_{d} \\ Y_{d} \end{array} \right)
\left( \begin{array}{c} d_L \\ d_L^\prime \end{array} \right),
\ee
where $X_{u,d}$ and $Y_{u,d}$ are $3 \times 6$, and CKM matrices are given by $V_{CKM} = X_u^\dagger X_d$ and $V_{CKM}^\prime = Y_u^\dagger Y_d$.

\begin{figure}[t!]
 \begin{center}
  \includegraphics[scale=0.5]{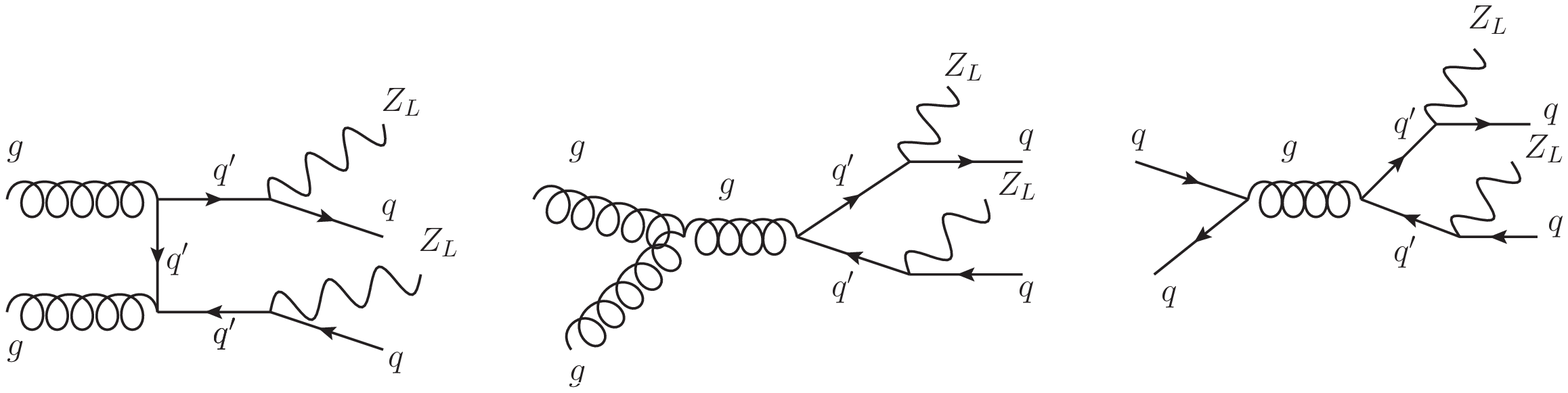}
    \caption{The pair production of new quarks at the LHC and their subsequent decay to the SM $Z_L$ boson and a quark.}
    \label{fig1}
     \end{center}
\end{figure}

Now we turn our attention towards phenomenological signatures and consequences of this model.  Due to mixing of the SM and new fermions, the charged and neutral current Lagrangians  allow new fermions $\psi^\prime$ to decay into a SM $W_L$ or $Z_L$ boson in association with a SM quark.  New fermions are charged under the colour gauge group $SU(3)_c$. Due to this, we can, for instance, produce new fermions $\psi^\prime$  via gluon-gluon and quark-antiquark initial states as shown in Fig.\ref{fig1} at the LHC.  

\begin{figure}[t!]
 \begin{center}
  \includegraphics[scale=0.7]{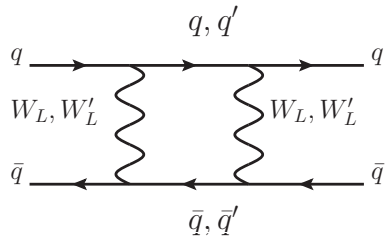}
    \caption{New box diagrams contributing to $K$ and $B$ meson mixing.}
    \label{fig2}
    \end{center}
\end{figure}

There are  flavour changing neutral current interactions at tree level in this model.  Hence, $K$ and $B$ mesons mixing, as shown in Fig.\ref{fig2},   is expected to place non-trivial constraints on the masses of the new gauge bosons  and fermions.  For instance,  diagrams which may play a non-trivial role are the ones which have a $W_L$ or $W_L^\prime$ with new fermions  in the box.

A simple UV completion of this model can come from vector-like iso-singlet quarks and leptons.  We observe that at least two vector-like isosinglet quarks of up and down type, one iso-singlet vector-like charged lepton, one iso-singlet vector-like neutrino, and their $CP$-counterparts are sufficient to provide UV completion of this model.  Their quantum numbers under $SU(3)_c \otimes SU(2)_L \otimes SU(2)_L^\prime  \otimes U(1)_{Y^\prime} $ and discrete symmetries $\mathcal{Z}_2$ and  $\mathcal{Z}_2^\prime$ given as follows:
\begin{eqnarray}
 Q= U_{L,R}:(3,1,1,\dfrac{4}{3})_{+-};D_{L,R}:(3,1,1,-\dfrac{2}{3})_{+-}, L= E_{L,R}:(1,1,1,-2)_{+-}; N_{L,R}:(1,1,1,0)_{+-},
\end{eqnarray}
and their $CP$-counter parts are,
\begin{eqnarray}
 Q^\prime= U_{L,R}^\prime:(3,1,1,\dfrac{4}{3})_{-+};D_{L,R}^\prime:(3,1,1,-\dfrac{2}{3})_{-+},L^\prime= E_{L,R}^\prime:(1,1,1,-2)_{-+}; N_{L,R}^\prime:(1,1,1,0)_{-+},
\end{eqnarray}
where charges of $\mathcal{Z}_2$ and  $\mathcal{Z}_2^\prime$ are given in sub-script.

\begin{figure}[t!]
 \begin{center}
  \includegraphics[scale=0.7]{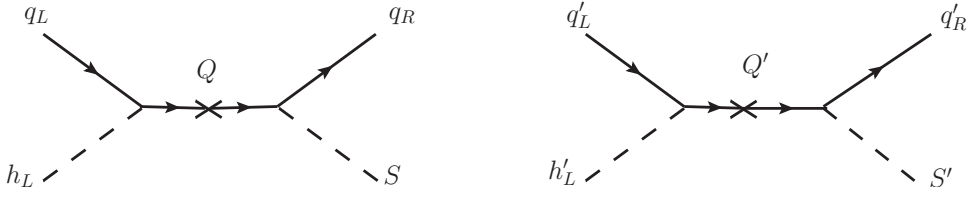}
    \caption{The explicit realization of the SM and new fermions $\psi^\prime$ in Eq.(\ref{mass1}) by the UV completion of the model.  Here, $q_L$, $q_R$  and $h_L$ are the SM quarks and Higgs boson, $q_L^\prime$, $q_R^\prime$  and $h_L^\prime$  are $CP$ counter parts,  $Q$ and $Q^\prime$ are  vector-like iso-singlet quarks, $S$ and $S^\prime$ are scalar particles corresponding to the scalar field $\chi$ and $\chi^\prime$.}
    \label{fig3}
    \end{center}
\end{figure}

The mass terms for vector-like fermions are the following:
\begin{eqnarray}
\label{vec1}
\mathcal{L}_{V} &=&\left( M_U \bar{U}_L U_R + M_U^* \bar{U}_L^\prime  U_R^\prime \right) + \left( M_D \bar{D}_L D_R + M_D^* \bar{D}_L^\prime  D_R^\prime \right) \\ \nonumber
&+&  \left(M_E  \bar{E}_L E_R  + M_E^* \bar{E}_L^\prime E_R^\prime \right) +  \left(M_N  \bar{N}_L N_R  + M_N^* \bar{N}_L^\prime N_R^\prime \right)+ {\rm H.c.}.
\end{eqnarray}
The interactions of the vector-like fermions with the SM and mirror fermions are given by,
\begin{eqnarray}
\label{vec2}
\mathcal{L}_{Vff^\prime}^\prime =  \left[ y^\prime  \bar{q}_L \varphi_L Q_R  + y^{\prime *} \bar{q}_L^\prime \varphi_L^\prime Q_R^\prime \right] +  \left[ c^\prime  \bar{l}_L \varphi_L L_R  +c^{\prime *} \bar{l}_L^\prime \varphi_L^\prime  L_R^\prime \right] + {\rm H.c}.
\end{eqnarray}
The interactions of singlet SM and mirror fermions with vector-like fermions are described by the following Lagrangian: 
\begin{eqnarray}
\label{vec3}
\mathcal{L}_{Vff^\prime}^{\prime \prime} =  \left[ y^{\prime \prime}  \bar{q}_R  Q_L \chi + y^{\prime \prime *} \bar{q}_R^\prime Q_L^\prime \chi^\prime \right] +  \left[c^{\prime \prime}  \bar{l}_R  L_L \chi + c^{\prime \prime *}  \bar{l}_R^\prime L_L^\prime \chi^\prime \right] + {\rm H.c}.
\end{eqnarray}
The Eqs.(\ref{vec1}), (\ref{vec2}) and (\ref{vec3}) provide a realization of the masses of the SM and mirror fermions given in Eq.(\ref{mass1}) as shown in Fig.\ref{fig3}.

Now we discuss consequences of this model for recently observed anomalies in flavour physics. The first deviation is observed in  $B \rightarrow K ll $ and  $B \rightarrow K^* ll$ decays which proceed through  $b \rightarrow s l^+ l^-$  transition. The optimised observable $P_5^\prime$ \cite{Descotes-Genon:2013vna}  is showing a deviation of $3.7 \sigma$ from the SM as measured by the LHCb  \cite{Aaij:2013qta}.  Moreover,  the ratio $R_K = \mathcal{B}_{B \rightarrow K \mu^+ \mu^-} /  \mathcal{B}_{B \rightarrow K e^+ e^-}$ measured by the LHCb is hinting towards  lepton flavour universality (LFU) violation \cite{Aaij:2014ora}.  Furthermore, the ratio $R_{K^*} = \mathcal{B}_{B \rightarrow K^* \mu^+ \mu^-} /  \mathcal{B}_{B \rightarrow K^* e^+ e^-}$ is recently measured by the LHCb, and this is also showing significant deviation from the SM prediction and lepton-flavour universality \cite{LHCb0417}.   Moreover, there is one more deviation from the SM expectation in the $b \rightarrow c l \nu$ transition having different final state leptons leading to the LFU violation in the observable $R_{D^*} =  \Gamma(B \rightarrow D^* \tau \nu) /  \Gamma(B \rightarrow D^* l \nu)$.  The most recent average of this observable is $4 \sigma$ away from the SM expectation \cite{Amhis:2014hma}. 

\begin{figure}[t!]
 \begin{center}
  \includegraphics[scale=0.7]{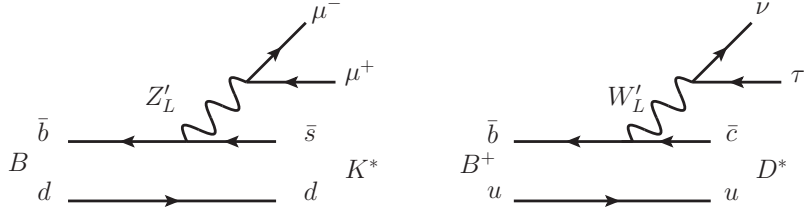}
    \caption{The contribution of new heavy gauge bosons to $B \rightarrow K^* \mu^+ \mu^-$ and $B^+ \rightarrow D^* \tau \nu$ at tree-level in the model discussed in this paper.}
    \label{fig4}
    \end{center}
\end{figure}

In this model,  LFU violation is introduced via  mixing of the SM fermions with new fermions,  and anomalies may be explained simultaneously due to a contribution coming from a new neutral and a charged gauge boson as shown in Fig.\ref{fig4}.   
Anomalies in $B \rightarrow K ll $ and  $B \rightarrow K^* ll$ decays may be explained via additional contribution of the $Z_L^\prime$ boson.  The ratio $R_{D^*} $ may be explained by observing couplings of $W_L^\prime$ to a charged lepton and a neutrino.  This will be investigated in a future study. 

It should be noted that this model is not constructed to explain above mentioned anomalies. The aim of this model is to restore $CP$ symmetry, which is a more fundamental discrete symmetry than $C$ or $P$  in the sense that it can distinguish matter and antimatter in an absolute and convention-independent way.  However, this model may explain above mentioned anomalies simultaneously which is not ad hoc at all.  The model also alleviates the fine tuning of the SM Higgs boson relatively in a small amount and provide possible dark matter candidates in the form of a real scalar particle (and the SM right-handed neutrinos and new neutrinos) corresponding to scalar field $\chi$ which is testable, for instance, in the process of gluon fusion $gg \rightarrow h_L S$ at the LHC.

\section*{Acknowledgements}
I thank to the anonymous reviewer for several important comments and suggestions.


\begin{thebibliography}{99}
\bibitem{Pati:1974yy} 
  J.~C.~Pati and A.~Salam,
  %``Lepton Number as the Fourth Color,''
  Phys.\ Rev.\ D {\bf 10}, 275 (1974)
  [Phys.\ Rev.\ D {\bf 11}, 703 (1975)].
  %%CITATION = PHRVA,D10,275;%%
  %3836 citations counted in INSPIRE as of 28 Sep 2015

\bibitem{Mohapatra:1974hk} 
  R.~N.~Mohapatra and J.~C.~Pati,
  %``Left-Right Gauge Symmetry and an Isoconjugate Model of CP Violation,''
  Phys.\ Rev.\ D {\bf 11}, 566 (1975).
  %%CITATION = PHRVA,D11,566;%%
  %1730 citations counted in INSPIRE as of 28 Sep 2015

\bibitem{Senjanovic:1975rk} 
  G.~Senjanovic and R.~N.~Mohapatra,
  %``Exact Left-Right Symmetry and Spontaneous Violation of Parity,''
  Phys.\ Rev.\ D {\bf 12}, 1502 (1975).
  %%CITATION = PHRVA,D12,1502;%%
  %1617 citations counted in INSPIRE as of 28 Sep 2015

%\cite{Mohapatra:1979ia}
\bibitem{Mohapatra:1979ia} 
  R.~N.~Mohapatra and G.~Senjanovic,
  %``Neutrino Mass and Spontaneous Parity Violation,''
  Phys.\ Rev.\ Lett.\  {\bf 44}, 912 (1980).
  doi:10.1103/PhysRevLett.44.912
  %%CITATION = doi:10.1103/PhysRevLett.44.912;%%
  %3843 citations counted in INSPIRE as of 05 Dec 2015


%\cite{Abbas:2016xgj}
\bibitem{Abbas:2016xgj} 
  G.~Abbas,
  %``Right-right-left extension of the standard model,''
  Mod.\ Phys.\ Lett.\ A {\bf 31}, no. 19, 1650117 (2016)
  doi:10.1142/S0217732316501170
  [arXiv:1605.02497 [hep-ph]].
  %%CITATION = doi:10.1142/S0217732316501170;%%
  %1 citations counted in INSPIRE as of 09 Aug 2016


%\cite{Abbas:2016qqc}
\bibitem{Abbas:2016qqc} 
  G.~Abbas,
  %``Low scale left-right-right-left symmetry,''
  Phys.\ Rev.\ D {\bf 95}, no. 1, 015029 (2017)
  doi:10.1103/PhysRevD.95.015029
  [arXiv:1609.02899 [hep-ph]].
  %%CITATION = doi:10.1103/PhysRevD.95.015029;%%
  %1 citations counted in INSPIRE as of 03 Mar 2017


 \bibitem{Foot:1991bp} 
  R.~Foot, H.~Lew and R.~R.~Volkas,
  %``A Model with fundamental improper space-time symmetries,''
  Phys.\ Lett.\ B {\bf 272}, 67 (1991).
  doi:10.1016/0370-2693(91)91013-L
  %%CITATION = doi:10.1016/0370-2693(91)91013-L;%%
  %297 citations counted in INSPIRE as of 07 Mar 2016

\bibitem{Silagadze:1995tr} 
  Z.~K.~Silagadze,
  %``Neutrino mass and the mirror universe,''
  Phys.\ Atom.\ Nucl.\  {\bf 60}, 272 (1997)
  [Yad.\ Fiz.\  {\bf 60N2}, 336 (1997)]
  [hep-ph/9503481].
  %%CITATION = HEP-PH/9503481;%%
  %99 citations counted in INSPIRE as of 07 Mar 2016




\bibitem{Berezhiani:1995yi} 
  Z.~G.~Berezhiani and R.~N.~Mohapatra,
  %``Reconciling present neutrino puzzles: Sterile neutrinos as mirror neutrinos,''
  Phys.\ Rev.\ D {\bf 52}, 6607 (1995)
  doi:10.1103/PhysRevD.52.6607
  [hep-ph/9505385].
  %%CITATION = doi:10.1103/PhysRevD.52.6607;%%
  %332 citations counted in INSPIRE as of 07 Mar 2016

%\cite{Cui:2011wk}
\bibitem{Cui:2011wk} 
  J.~W.~Cui, H.~J.~He, L.~C.~Lu and F.~R.~Yin,
  %``Spontaneous Mirror Parity Violation, Common Origin of Matter and Dark Matter, and the LHC Signatures,''
  Phys.\ Rev.\ D {\bf 85}, 096003 (2012)
  doi:10.1103/PhysRevD.85.096003
  [arXiv:1110.6893 [hep-ph]].
  %%CITATION = doi:10.1103/PhysRevD.85.096003;%%
  %40 citations counted in INSPIRE as of 06 Jun 2017



%\cite{Cui:2012mq}
\bibitem{Cui:2012mq} 
  J.~W.~Cui, H.~J.~He, L.~C.~Lü and F.~R.~Yin,
  %``GeV Scale Asymmetric Dark Matter from Mirror Universe: Direct Detection and LHC Signatures,''
  Int.\ J.\ Mod.\ Phys.\ Conf.\ Ser.\  {\bf 10}, 21 (2012)
  doi:10.1142/S2010194512005727
  [arXiv:1203.0968 [hep-ph]].
  %%CITATION = doi:10.1142/S2010194512005727;%%
  %4 citations counted in INSPIRE as of 06 Jun 2017



%\cite{Gu:2014mga}
\bibitem{Gu:2014mga} 
  P.~H.~Gu,
  %``A new leptogenesis scenario parametrized by Dirac neutrino mass matrix,''
  arXiv:1410.5753 [hep-ph].
  %%CITATION = ARXIV:1410.5753;%%
  %2 citations counted in INSPIRE as of 19 Jul 2017

%\cite{Gu:2017mkm}
\bibitem{Gu:2017mkm} 
  P.~H.~Gu,
  %``Spontaneous mirror left-right symmetry breaking for leptogenesis parametrized by Majorana neutrino mass matrix,''
  arXiv:1706.07706 [hep-ph].
  %%CITATION = ARXIV:1706.07706;%%






%\cite{Lee:1973iz}
\bibitem{Lee:1973iz} 
  T.~D.~Lee,
  %``A Theory of Spontaneous T Violation,''
  Phys.\ Rev.\ D {\bf 8}, 1226 (1973).
  doi:10.1103/PhysRevD.8.1226
  %%CITATION = doi:10.1103/PhysRevD.8.1226;%%
  %1065 citations counted in INSPIRE as of 02 Jun 2017

%\cite{Branco:1980sz}
\bibitem{Branco:1980sz} 
  G.~C.~Branco,
  %``Spontaneous {CP} Nonconservation and Natural Flavor Conservation: A Minimal Model,''
  Phys.\ Rev.\ D {\bf 22}, 2901 (1980).
  doi:10.1103/PhysRevD.22.2901
  %%CITATION = doi:10.1103/PhysRevD.22.2901;%%
  %144 citations counted in INSPIRE as of 02 Jun 2017


%\cite{Deshpande:1976yp}
\bibitem{Deshpande:1976yp} 
  N.~G.~Deshpande and E.~Ma,
  %``Comment on Weinberg's Gauge Theory of CP Nonconservation,''
  Phys.\ Rev.\ D {\bf 16}, 1583 (1977).
  doi:10.1103/PhysRevD.16.1583
  %%CITATION = doi:10.1103/PhysRevD.16.1583;%%
  %110 citations counted in INSPIRE as of 02 Jun 2017


%\cite{Lavoura:1994fv}
\bibitem{Lavoura:1994fv} 
  L.~Lavoura and J.~P.~Silva,
  %``Fundamental CP violating quantities in a SU(2) x U(1) model with many Higgs doublets,''
  Phys.\ Rev.\ D {\bf 50}, 4619 (1994)
  doi:10.1103/PhysRevD.50.4619
  [hep-ph/9404276].
  %%CITATION = doi:10.1103/PhysRevD.50.4619;%%
  %118 citations counted in INSPIRE as of 02 Jun 2017



%\cite{Barger:1996jc}
\bibitem{Barger:1996jc} 
  V.~D.~Barger, A.~K.~Das and C.~Kao,
  %``The Electric dipole moment of the muon in a two - Higgs doublet model,''
  Phys.\ Rev.\ D {\bf 55}, 7099 (1997)
  doi:10.1103/PhysRevD.55.7099
  [hep-ph/9611344].
  %%CITATION = doi:10.1103/PhysRevD.55.7099;%%
  %30 citations counted in INSPIRE as of 02 Jun 2017



%\cite{Lavoura:1997pq}
\bibitem{Lavoura:1997pq} 
  L.~Lavoura,
  %``A New type of spontaneous CP breaking,''
  Phys.\ Lett.\ B {\bf 400}, 152 (1997)
  doi:10.1016/S0370-2693(97)00344-4
  [hep-ph/9701221].
  %%CITATION = doi:10.1016/S0370-2693(97)00344-4;%%
  %10 citations counted in INSPIRE as of 02 Jun 2017



%\cite{Chakdar:2013tca}
\bibitem{Chakdar:2013tca} 
  S.~Chakdar, K.~Ghosh, S.~Nandi and S.~K.~Rai,
  %``Collider signatures of mirror fermions in the framework of a left-right mirror model,''
  Phys.\ Rev.\ D {\bf 88}, no. 9, 095005 (2013)
  doi:10.1103/PhysRevD.88.095005
  [arXiv:1305.2641 [hep-ph]].
  %%CITATION = doi:10.1103/PhysRevD.88.095005;%%
  %7 citations counted in INSPIRE as of 07 Mar 2016



\bibitem{Gu:2012in} 
  P.~H.~Gu,
  %``Mirror left-right symmetry,''
  Phys.\ Lett.\ B {\bf 713}, 485 (2012)
  doi:10.1016/j.physletb.2012.06.042
  [arXiv:1201.3551 [hep-ph]].
  %%CITATION = doi:10.1016/j.physletb.2012.06.042;%%
  %7 citations counted in INSPIRE as of 07 Mar 2016




%\cite{Abbas:2017hzw}
\bibitem{Abbas:2017hzw} 
  G.~Abbas,
  %``A low scale left-right symmetric mirror model,''
  arXiv:1706.01052 [hep-ph].
  %%CITATION = ARXIV:1706.01052;%%



%\cite{Schmaltz:2002wx}
\bibitem{Schmaltz:2002wx} 
  M.~Schmaltz,
  %``Physics beyond the standard model (theory): Introducing the little Higgs,''
  Nucl.\ Phys.\ Proc.\ Suppl.\  {\bf 117}, 40 (2003)
  doi:10.1016/S0920-5632(03)01409-9
  [hep-ph/0210415].
  %%CITATION = doi:10.1016/S0920-5632(03)01409-9;%%
  %167 citations counted in INSPIRE as of 21 Jun 2017




%\cite{Aaij:2012nna}
\bibitem{Aaij:2012nna} 
  R.~Aaij {\it et al.} [LHCb Collaboration],
  %``First Evidence for the Decay $B_s^0 \to \mu^+ \mu^-$,''
  Phys.\ Rev.\ Lett.\  {\bf 110}, no. 2, 021801 (2013)
  doi:10.1103/PhysRevLett.110.021801
  [arXiv:1211.2674 [hep-ex]].
  %%CITATION = doi:10.1103/PhysRevLett.110.021801;%%
  %434 citations counted in INSPIRE as of 16 Jun 2017


%\cite{CMS:2014xfa}
\bibitem{CMS:2014xfa} 
  V.~Khachatryan {\it et al.} [CMS and LHCb Collaborations],
  %``Observation of the rare $B^0_s\to\mu^+\mu^-$ decay from the combined analysis of CMS and LHCb data,''
  Nature {\bf 522}, 68 (2015)
  doi:10.1038/nature14474
  [arXiv:1411.4413 [hep-ex]].
  %%CITATION = doi:10.1038/nature14474;%%
  %302 citations counted in INSPIRE as of 16 Jun 2017





%\cite{Descotes-Genon:2013vna}
\bibitem{Descotes-Genon:2013vna}
  S.~Descotes-Genon, T.~Hurth, J.~Matias and J.~Virto,
  %``Optimizing the basis of $B\to K^*ll$ observables in the full kinematic range,''
  JHEP {\bf 1305} (2013) 137
  doi:10.1007/JHEP05(2013)137
  [arXiv:1303.5794 [hep-ph]].
  %%CITATION = doi:10.1007/JHEP05(2013)137;%%
  %184 citations counted in INSPIRE as of 18 May 2017


%\cite{Aaij:2013qta}
\bibitem{Aaij:2013qta} 
  R.~Aaij {\it et al.} [LHCb Collaboration],
  %``Measurement of Form-Factor-Independent Observables in the Decay $B^{0} \to K^{*0} \mu^+ \mu^-$,''
  Phys.\ Rev.\ Lett.\  {\bf 111}, 191801 (2013)
  doi:10.1103/PhysRevLett.111.191801
  [arXiv:1308.1707 [hep-ex]].
  %%CITATION = doi:10.1103/PhysRevLett.111.191801;%%
  %329 citations counted in INSPIRE as of 18 May 2017



%\cite{Aaij:2014ora}
\bibitem{Aaij:2014ora} 
  R.~Aaij {\it et al.} [LHCb Collaboration],
  %``Test of lepton universality using $B^{+}\rightarrow K^{+}\ell^{+}\ell^{-}$ decays,''
  Phys.\ Rev.\ Lett.\  {\bf 113}, 151601 (2014)
  doi:10.1103/PhysRevLett.113.151601
  [arXiv:1406.6482 [hep-ex]].
  %%CITATION = doi:10.1103/PhysRevLett.113.151601;%%
  %357 citations counted in INSPIRE as of 18 May 2017

\bibitem{LHCb0417}
S. Bifani. Search for new physics with b s decays at LHCb - 2017. (on behalf of the LHCb
collaboration) seminar presented at CERN on 18 Apr. 


%\cite{Amhis:2014hma}
\bibitem{Amhis:2014hma} 
  Y.~Amhis {\it et al.} [Heavy Flavor Averaging Group (HFAG)],
  %``Averages of $b$-hadron, $c$-hadron, and $\tau$-lepton properties as of summer 2014,''
  arXiv:1412.7515 [hep-ex].
  %%CITATION = ARXIV:1412.7515;%%
  %454 citations counted in INSPIRE as of 18 May 2017
  
  




 \end{thebibliography}
\end{document}